**Modeling the dynamics of natural rotifer populations:**

**phase-parametric analysis**

Faina Berezovskaya [1*], Georgy Karev [3], and Terry W. Snell [2]

[1] Department of Mathematics, Howard University, Washington, DC 20059

[2] School of Biology, Georgia Institute of Technology, Atlanta, GA 30332-0145

[3] Oak Ridge Institute for Science and Education, National Institute of Health, Bethesda, MD 20894

 *Corresponding author

**Running Head:  Modeling natural rotifer populations**

**Corresponding author:** Faina S. Berezovskaya,

Department of Mathematics, Howard University, Washington, DC 20059

**E-mail: fsberezo@hotmail.com, FAX:202-806-6831**

**Abstract**

A model of the dynamics of natural rotifer populations is described as a discrete nonlinear map depending on three parameters, which reflect characteristics of the population and environment. Model dynamics and their change by variation of these parameters were investigated by methods of bifurcation theory.  A phase–parametric portrait of the model was constructed and domains of population persistence (stable equilibrium, periodic and a- periodic oscillations of population size) as well as population extinction were identified and investigated. The criteria for population persistence and approaches to determining critical parameter values are described. The results identify parameter values that lead to population extinction under various environmental conditions. They further illustrate that the likelihood of extinction can be substantially increased by small changes in environmental quality, which shifts populations into new dynamical regimes.

## 1.  Introduction

Because of the ecological importance of zooplankton, a substantial amount of data exists on the abundance of natural populations in lakes, estuaries and coastal marine environments. A wide variety of mathematical models have been employed in modeling the dynamics of natural zooplankton populations (McCauley et al. 1996, Snell and Serra, 1998). However, systematic study of the dynamical behavior of these models through all phase space is lacking. In this paper, we study mathematical models developed to analyze the dynamics of natural rotifer populations and to evaluate the ecological effects of toxicant exposure (Snell and Serra, 1998, 2000). The models are based on the supposition that the dynamics of natural populations constitute a complex mixture of deterministic and chaotic components and are inherently non-linear (Turchin and Taylor, 1992, Berezovskaya and Karev, 1998). However, it is well known that steady states, complex oscillations and even stochastic dynamical regimes may happen and be stable in deterministic models. These phenomena were discovered through the analysis of the classical population map models (Shapiro, 1972, May, 1975, et al.).

New methods have been developed to extract deterministic dynamics components from short noisy ecological time series data (Royama, 1992, Ellner and Turchin, 1995). These methods in (Snell and Serra, 1998) were applied to data from natural populations of nine rotifer species *Asplanchna girodi, Filinia pejleri, Keratella tropica, Monostylla bulla, Brachionus rotundiformis* and four others belonging to the genus *Brachionus.* Time series of a population density $N(t)$ (a number of organisms per liter at time $t$ with time unit equal to 2 days) have been received. Using these series the RAMAS–program from Applied Biomathematics (Aksakaya et. al., 1999) permitted construction of seven phenomenological mathematical models of population dynamics. These models represent maps

$$N(t+1) = G(N(t)) = N(t)R(N(t)) \tag{1}$$

where $N>0$, $G(0)=0$. *Ricker, logistic, Hassel model*s and others (see, Thunberg, 2000, Devaney, 1998, Blokh, Lyubich, 1991, etc.) have been checked for fitting.

The best-fit model for 5 of 9 data sets was named the Consensus model. It has $R(N(t))= \exp\{r(N(t))\}$ and thus takes the form:

$$N(t+1) = N(t)\exp\{r(N(t))\}, \tag{2}$$

where $r(N)$ is a growth rate of a population (in logarithmic scale):

$$r(N) = -a + b/N - c/N^2. \tag{3}$$

with constants *a, b, c* distinguished for every of the nine rotifer species; these values are given in the Table 1 (Snell and Serra, 1998) and in Table 1 below. Remark, that *a* and *b* are positive for any indicated species, while *c* can be both positive and negative. Figures 1a-e illustrate the dynamical regimes of the Consensus model; these regimes arise just for values taken from Table 1.

Fig. 1a shows phase trajectories of population growth for the rotifer *Asplanchna girodi*. The model has three equilibria, two of which are stable; their domains of attraction are separated by the unstable equilibrium. Dependently on initial values, a trajectory monotonically approaches

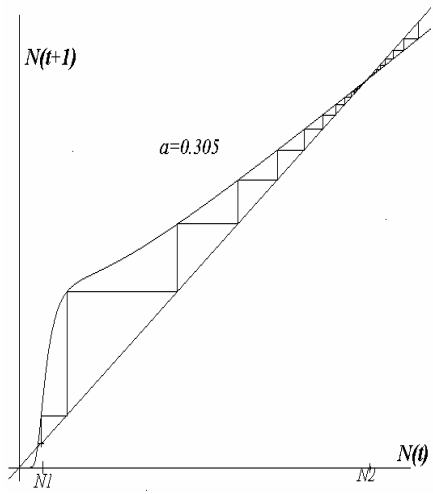

**a**

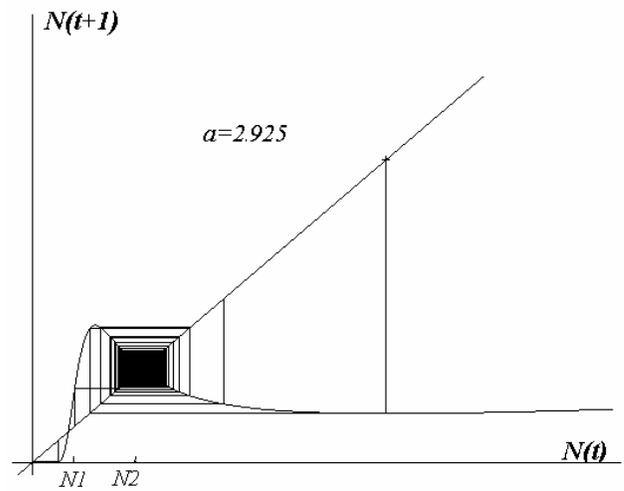

**b**

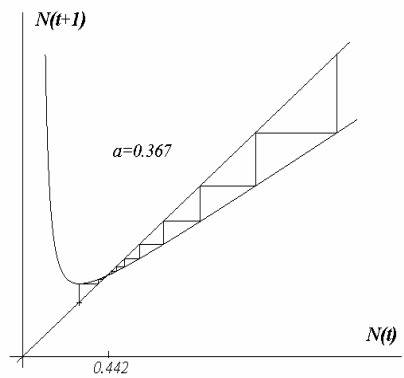

**c**

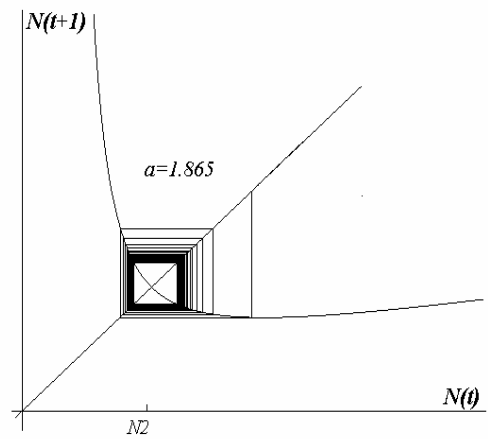

**d**

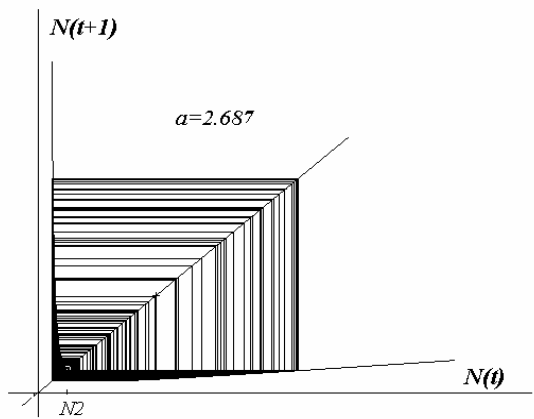

**e**

**Fig. 1**. Phase portraits of Consensus model (2), (5):

**a** – *Asplanchna girodi*; two 'nontrivial' equilibria: unstable $N_1$=0.013, and stable $N_2$=0.253;

**b** – *B.budapestinensis*; two 'nontrivial' equilibria: unstable $N_1$=0.294 and stable $N_2$=0.912;

**c** - *Monostylla bulla*; unstable nontrivial equilibrium $N_2$=0.834 and 2-periodic stable oscillations;

**d** - *B.angularis*; 'nontrivial' equilibrium $N_2$ =1.511 and stable many/non-periodic oscillations;

**e** – *Filinia pejleri*; stable 'nontrivial' equilibrium $N_2$ = 0.442.

one of the stable equilibria. Remark that this bistable dynamics is intrinsic also to the models of the rotifers: *Keratella tropica*, *Brachionus dichotomus*, *B. liratus.*

The model of *B.budapestinensis* (Fig.1b) shows oscillations with decreasing amplitude when a trajectory is approaching a non-zero equilibrium. Similar type of dynamics is intrinsic to the model of *B.rotundiformis* populations.

The model of *Filinia pejleri* (Fig.1c) has only one non-zero stable equilibrium and approaches it from any initial state out of the origin. Trajectories are monotone if the initial values are close to the equilibrium and are similar to those in Fig.1a.

The *Monostylla bulla* population undergoes "simple" 2-periodic oscillations as stable dynamic regime (Fig. 1d).

Dynamics of population of *B.angularis* has many-periodic or a-periodic oscillations as a stable dynamic regime (Fig.1e). This regime is realized for any non-zero initial population.

Thus, the Consensus model is able to describe a wide range of dynamical behavior intrinsic to different rotifer species in various environmental conditions: steady state, periodic oscillations and near chaotic oscillations. We emphasize that these dynamical regimes pertain to real data from natural rotifer populations.

We further show that with change of coefficients the model demonstrates practically all types of non-linear dynamics discovered in classical models (Shapiro, 1972; May, 1975). Some types of the model dynamics illustrated in Figs 1a-e are similar to those observed in *Ricker, logistic, Hassel* models. Moreover, the Consensus model possesses the important property of bistability (lacking in the models mentioned above); due to this property the model realizes either steady state (persistence or extinction), whose attracting basins are divided by an unstable fixed point.

## 2. Statement of the problem, re-parameterization, description of results

Suppose $a>0$, $b>0$, $c$ be parameters of the model (2),(3). Remark, that $r(N)$ in (3) consists of two *independent* parts:

(i)     density dependent component:  $b/N – c/N^2$,

(ii)     density independent component:  $– a$.

Thus, parameters $b$ and $c$ can be related to population-specific characteristics, like the response to increased population density, while parameter $a$ characterizes density-independent effects on birth and death rates such as poor water quality, extreme temperature, or toxicant exposure; growth of $a$ leads to decreasing of $r(N)$ so it could be interpreted as an environmental press to rotifers.

The number of parameters in equations (2), (3) can be reduced to two. Setting

$$N \rightarrow bN , \quad \gamma = c/b^2, \qquad (4)$$

one has

$$r(N) = -a + 1/N – \gamma/N^2. \qquad (5)$$

Scaling (4) does not change "environmental density-independent" parameter $a$, however it introduces "cumulative density-dependent" parameter $\gamma$, which we will call $\gamma$-*index* of a population. The Table 1 contains values $\gamma$ defined by (4) and ordered by their $\gamma$-*index* the rotifer species**: 1.** *Keratella tropica*, **2**. *Asplanchna girodi,* **3.** *Brachionus dichotomus,* **4.** *B. liratus,* **5.** *B. budapestinensis,* **6.** *B. rotundiformis,* **7.** *Monostylla bulla,* **8.** *B. angularis,* **9.** *Filinia pejleri.* Thus, for given value of parameter $a$ the "typical" phase portraits of model species **1-4** are presented in Fig.1a and species **5, 6** in Fig.1b, the phase portrait of **7** is given in Fig.1d and **8** in Fig.1e; at last, Fig. 1c presents the phase portrait of model **9.**

In this work we analyze the model (1) of the form (2), (5) in parameter space $a$, $\gamma$ by methods of bifurcation theory. We show that dependently on parameters the model demonstrates a wide range of dynamics: equilibrium, oscillatory, chaotic. We define areas in parameter space corresponding to certain types of the model dynamics (see Fig.3 below) and indicate the boundaries of these areas. A population approaching the boundary may be in danger of extinction. Studies of the effects of toxicants on population dynamics have focused on direct effects on birth and death rates. In this paper, we show that toxicant exposure can shift the dynamical regime of population, indirectly increasing the likelihood of extinction. Bifurcation analysis allows us to evaluate toxicant effects by studying model dynamics with changing parameter values. The parameter points ($a$, $\gamma$) taken from Table 1 are considering as experimental data; they are presented in Fig.3 as numbered circles. Further we trace the behavior of respective models of species with increasing of toxicant press *a* (Fig.-s 6a, b, c).

### 3. Critical points and equilibrium points

The diversity of behaviors of the model is associated with the complex form of function $G(N)$ changing under change of parameters. The shape of $G(N)$ essentially depends on value $\gamma$. Coordinates of critical points satisfy the equation

$$G_N(N) \equiv \exp\{r(N))(1- 1/N+ 2\gamma/N^2 )\} =0, \tag{6}$$

whose analysis leads to the following statement.

*Proposition 1. The map (2),(5) for $N \geq 0, G \geq 0$*

(i)     *has no critical points for $\gamma > 1/8$,*

(ii)    *has two critical points: minimum at $N^+ = (1 + \sqrt{(1-8\gamma)} )/2$    and maximum at*

   $N^- = (1 - \sqrt{(1-8\gamma)} )/2$ , *for $0 < \gamma < 1/8$;*

(iii)   *has unique minimum at $N^+$ for $\gamma \leq 0$.*

(iv)    *has negative Schwarzian derivative $SG(N)$ for $\gamma < 1/8$.*

Proof of the Proposition is given in Appendix 1.

Three types of $G(N)$ are presented in Fig.-s 2a,b,c.

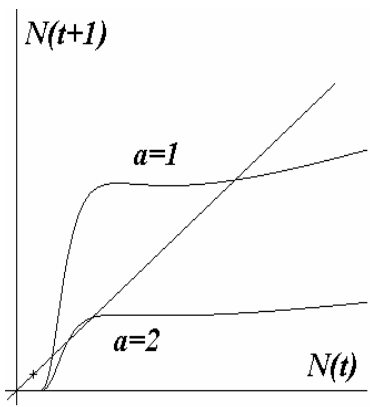

**a**

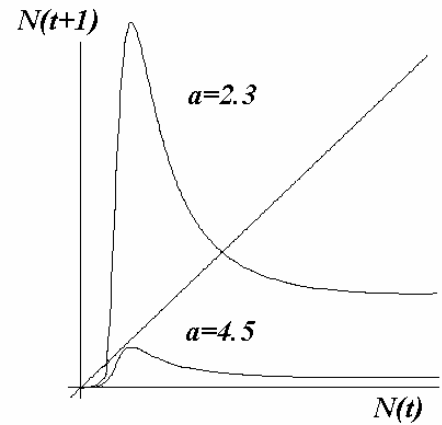

**b**

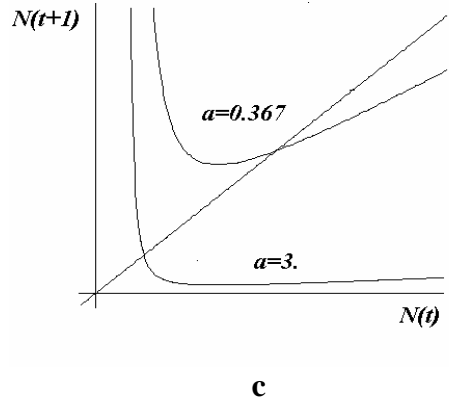

**c**

**Fig. 2.** The relationship between population growth rate $G(N)$ and density $N$ for model (1), (2): **a -** for c>0, $b^2$<8c, **b -** for $b^2$>8c>0, **c-** for $c$ <0 .

Fig.-s 2a and 2b correspond to a positive values of parameter $\gamma$; the case $\gamma$ >1/8 is shown in Fig. 2a  and 0<$\gamma$ < 1/8 in Fig. 2b. Fig. 2c corresponds to a negative γ. For γ >1/8 the function $G(N)$ monotonically increase (Fig.2a)*,* for 0<$\gamma$ <1/8 $G(N)$ is a function with "hump and tail" (Franke, Yakubu, 1994), see Fig. 2b, at last  for $\gamma$<0 $G(N)$ is a parabolic-like function with a single minimum (Fig.2c).  In all cases $G(N)$ has an asymptote $N$exp(-$a$), such that $G(N)/N$ <1 for any $a$>0, γ and large $N$.

Equilibria of model (2), (5) satisfy the equation

$$G(N) = N \qquad (7)$$

where

$$G(N) = N\exp(r(N)), \ r(N) = -a + 1/N - \gamma/N^2 \ . \qquad (8)$$

Equation (8) has the root $N_1$ =0 for any parameter value γ. Other roots of (8) can be obtained from the equation $r(N)$=0. Therefore, all non-trivial fixed points satisfy the equation

$$-a + 1/N - \gamma/N^2 = 0. \qquad (9)$$

The stability of a fixed point $N^*$ of a map $G(N)$ is defined by the value of the multiplier μ = $G_N(N^*)$ = $(1 - 1/N^* + 2\gamma/N^{*2})$exp($r(N^*)$). The point is stable if $|$μ$|$ <1 and unstable if $|$μ$|$>1 (see, for example, Devaney, 1989). Analysis of number of roots of (7), (8) and multipliers of $G$ in fixed points (see Fig.-s 2a-c) results as following statement.

*Proposition* 2. *(1) Model* (2)*,* (5) *has non-negative equilibria:*

*(1i) only N=0 for* $\gamma$ >1/(4a);

*(1ii) N=0, N=$N_1$, N=$N_2$  for* 0 <$\gamma$ <1/(4a); *N=0, N=$N_2$ for* $\gamma \le$ 0,

where $N_1 = (1 - \sqrt{(1-4a\gamma)}\,)/2a$,  $N_2 = (1 + \sqrt{(1-4a\gamma)}\,)/2a$, if $1-4a\gamma \geq 0$.

(2i) $N=0$ is stable if $\gamma >0$ and unstable if $\gamma \leq 0$,

(2ii) $N_1$ is unstable if $0 <\gamma <1/(4a)$,

(2iii) $N_2$ is stable for $0<a<1$ and  $\gamma <1/(4a)$ as well as for $a>1$ and

$(a-2)/(4(a-1)^2)<\gamma <1/(4a)$,  $N_2$ is unstable for $a>1$ and $\gamma <(a-2)/(4(a-1)^2)$.

Proof of the Proposition is given in Appendix 1.

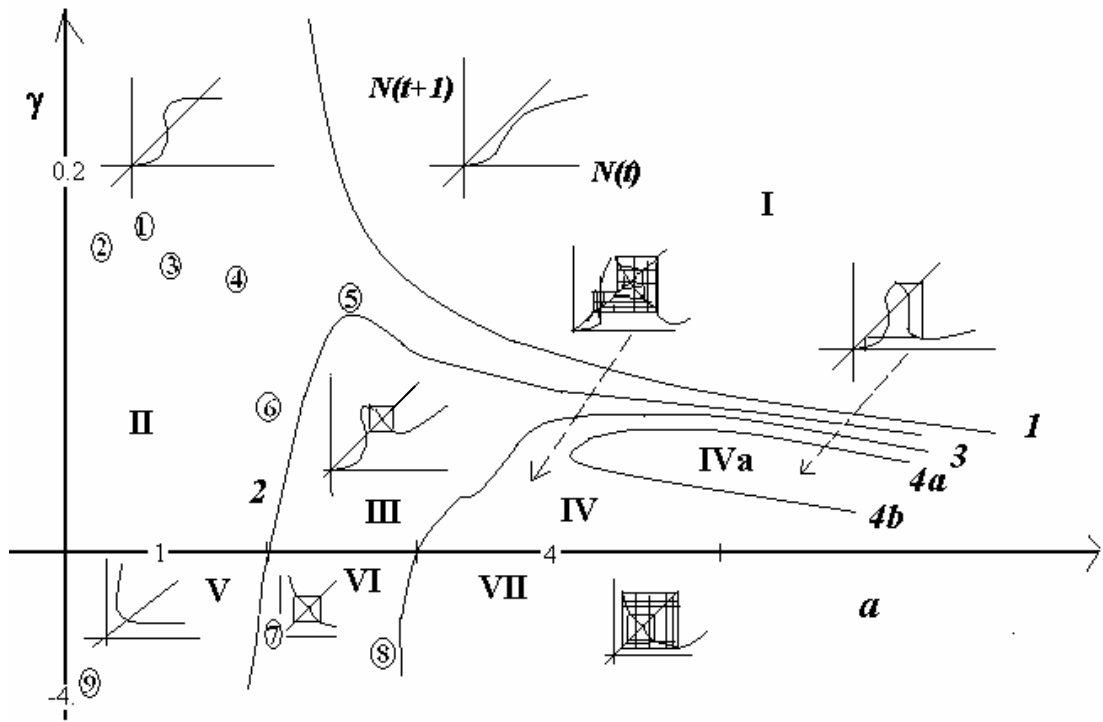

**Fig. 3.** Parameter-phase portrait of model (2), (5) in the half-plane $\{a>0, \gamma \}$ consists of eight domains:

I – Total Extinction, II – Bistability,  III – Zero stable and Periodic orbits,

IV – Oscillations and Chaos with Zero stable, IVa - Total Extinction in a-periodic oscillations

V – Monostability,  VI - Periodic orbits,  VII – Oscillations and Chaos Zero unstable.

The boundaries of the domains: $\gamma =0$, $\{\boldsymbol{1}\mid \gamma =1/4a\}$, $\{\boldsymbol{2}\mid \gamma =(a-2)/(4(a-1)^2\,), a>1\}$,

$\{\boldsymbol{3}\mid \gamma =\gamma*(a)$ is given in Tabl.2$\}$, $\{\boldsymbol{4}\mid \gamma =\gamma^y(a)$,  $\gamma =\gamma_d(a)$ for $a \geq a*=3.9388756$ are given in Tabl.2$\}$

Given rotifer species ordered by decrease of $\gamma$ is labeled by integer in a circle.

Due to Proposition 2, three parameter boundaries are defined at the phase -parameter portrait of the model (see Fig.3):

*line* $\gamma = 0$

dividing domains $\{a>0,\ \gamma>0\}$ where the model has from one up to three equilibria in the positive quadrant of $(N,G)$, and $\{a>0,\ \gamma<0\}$ where $G(N)$ has up to two equilibria. Changing of the stability of the equilibrium $N=0$ also happens at this line;

*line 1:* $\gamma = 1/(4a)$

dividing domains $\{a>0,\ \gamma>0\}$ into domain **I:** $\{a>0,\ \gamma>1/(4a)\}$ where the model has the only equilibrium $N=0$ and domain **II** $\{a>0,\ 0<\gamma<1/(4a)\}$ where the model has three equilibria $N=0$, $N_1$, $N_2$ (see Fig.3). With parameters belonging to the curve **1,** (9) has two-multiple root $N_0 = 1/(2a)$), and multiplier $G_N(N_0)=1$;

*line 2:* $\gamma = (a\text{-}2)/(4(a\text{-}1)^2)$

dividing domains $\{a>0,\ \gamma\}$ into parts domains **II** and **III** for $\gamma>0$ as well as **V** and **VI** for $\gamma<0$: equilibrium $N=N_2$ is stable in **II**, **V** and unstable in **III, VI** (see Fig.3). With parameter values $(\gamma*,\ a*)$ belonging to line **2**, the graph of $G(N;\gamma*,a*)$ crosses over the bisector $G=N$ in the point $N_2$ and $G^2(N;\gamma*,a*)$ cubically touches the bisector $G^2 = N$ in the point $N_2$. The multiplier $G_N(N_2)=\text{-}1$. We discuss the model behavior with crossing over line **2** below.

**Remark.** Let us emphasize a fundamental difference between instances with $\gamma>0$ (Fig.-s 2a, 2b) and $\gamma<0$ (Fig.2c). In both cases, $N=0$ is a fixed point of the map $G(N)$. For $\gamma>0$ the map is continuous at $N=0$ while it is the point of discontinuity of $G$ for $\gamma<0$. It is clear that the model with negative $\gamma$ for small $N$ seems to be biologically questionable because it represents very high growth rate of a population for small values of $N$, moreover, according to this model the population can not go extinct. This means that the model with $\gamma<0$ is reliable only for values of $N$ outside a vicinity of the origin.

### 4. Limit cycles and chaotic trajectories

It is known (see, for example, Devaney, 1989) that if a map $G(N)$ has a cycle $(N_1,...\ N_n)$ of period $n$, then each $N_i,\ i=1,...n$ is a fixed point of the map $G^n(N)=G(G...(G(N))..)$ obtained by $n$-multiple iterations of $G(N)$, this means that $G^n(N_i)=N_i$. A cycle $(N_1,...\ N_n)$ of period $n$ is attracting (or stable) if the product $\lvert G_N(N_1)...\ G_N(N_n)\rvert <1$.

The loss of stability of the point $N_2$ situated at the boundary **2** results in appearance 2-point cycle at the plane $(N,G(N))$. This cycle consists from the fixed points $N^1$, $N^2$ of the map $G^2(N)$, $N^1 < N_2 < N^2$, which satisfy the equation

$$N\exp(r(N))\exp(r(N\exp(r(N))) = N \qquad (10)$$

where $r(N)$ is defined in (8). Each non-zero root of (10) satisfies the equation

$$r(N) + r(N\exp(r(N)) = 0. \qquad (11)$$

Analysis of equations (7), (8), (11) (see Appendix 1) allows formulating

*Proposition 3.The flip bifurcation accompanied by the appearance of stable 2-point cycle is realized in the model* (**2**), (**5**) *when a parametric point* (*a, γ*) *crosses over the curve* **2**: $\gamma = (a-2)/(4(a-1)^2)$ (*from the top to the down*) *and enters the domains* **III** (*for γ≥0*) *or* **VI** (*for γ<0*), *see Fig.3.*

We will refer to curve **2** as the *flip* boundary $\gamma = \tilde{\gamma}(a)$. It is easy to see from the equation of line **2** that $\gamma = \tilde{\gamma}(a)$ has a maximum equal to $\gamma = 1/16$ in the point $a=3$. Note, that $\tilde{\gamma}(a)$ has the vertical asymptote at $a = 1$.

The evolution of a two-periodic cycle, which appears in the flip bifurcation, can be traced further with variation of parameters (Devaney, 1989, Kuznetsov, 1995, Thunberg, 2000). We studied the *Feigenbaum*`s cascade in model (2), (5), resulting the appearance of $2^n$–periodic stable cycle in $n^{th}$ iteration step, and verified it's existence numerically with the help of program "Dynamics–2" (Nusse & Yorke, 1997) for fixed γ in areas **III**, **IV** and **VI** (see Fig.3). Bifurcation diagrams are given in Figs. **4**a, b in coordinates (*parameter a, population size W(a)*), where $W(a)$ represents the set of all stable attractors of the map.

A stable many-periodic trajectory looks like a-periodic for an observer, so it is important to find the domain in parametric space, where the map (2), (5) shows really chaotic behavior. According to the Sarkovskii` theorem (Sarkovskii, 1964, Li & Yorke, 1975), the existence of a cycle of period 3 implies the existence of cycles of *any* period as well as a-periodic trajectories - "cycle period 3 introduces chaos".

The appearance of a *stable* 3-periodic cycle in the plane $(N,G)$ defines the parameter boundary **3** of the domain of "chaoticity". We calculated this boundary as follows. For a fixed γ the maximum $N(\gamma)$ of $G$ was taken as one of the points of the 3-cycle. Two other points correspond to the points of touching of $G^3(N)$ to the bisector (see Fig.13.1 from Devaney, 1989). The parameter curve **3**: γ=γ*(*a*), given in Table 1, was computed by the TRAX software program

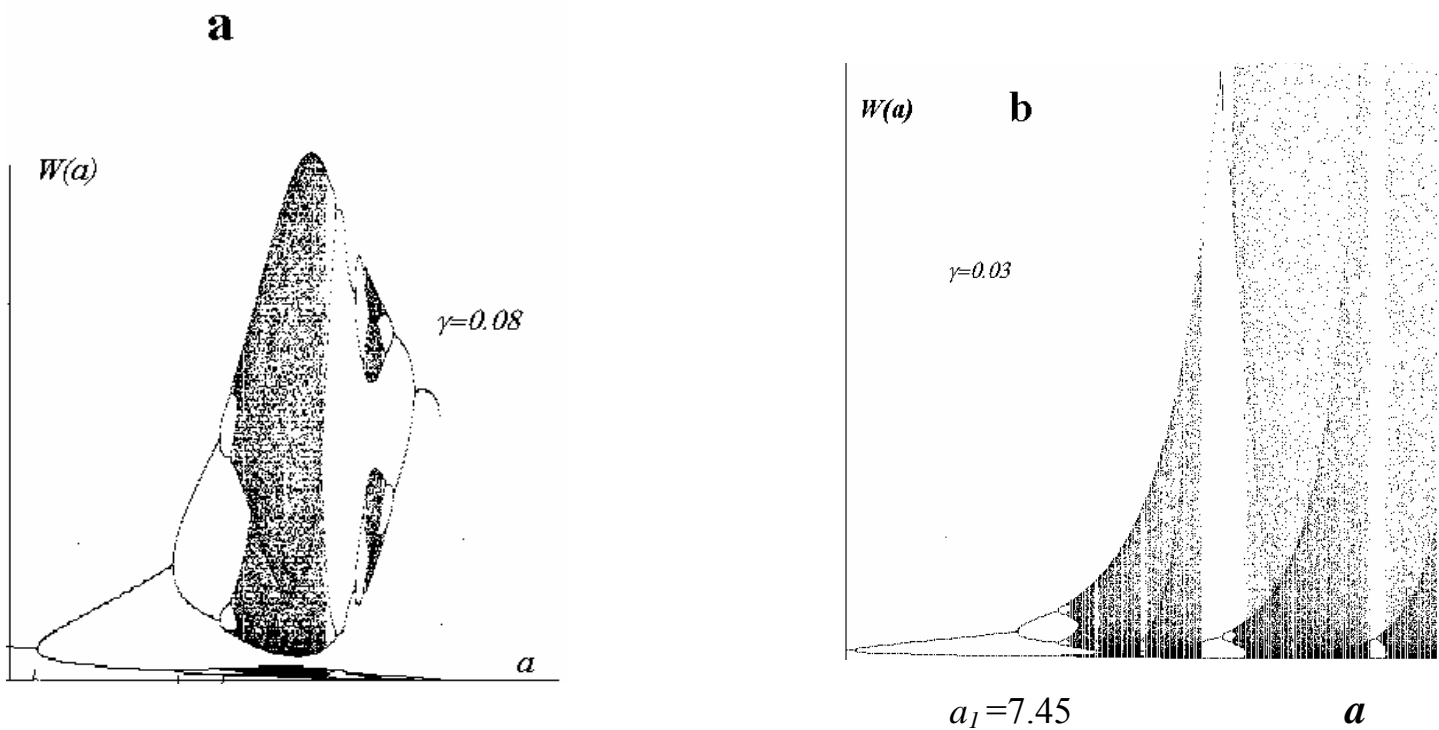

**Fig.4. a-** Bifurcation diagram in the domain III with γ=0.08 ($a_1$=2.163, $a_2$=2.889, $a_3$=3.137);
**b-**Fragment of bifurcation diagram in the domain IV (γ=0.03).

(Levitin, 1989). We refer to this curve as Sarkovskii`, or 3-period boundary. Curve **3** separates domains **III** and **IV** for γ >0 and **VI** and **VII** for γ<0 (see Fig.3). The multiplier of the considering 3-cycle equals to zero (because $N^-$ is on of the points of 3-cycle for any γ). Hence, every 3-period cycle, corresponding to the Sarkovskii boundary is attracting, see Fig.4b (but its attraction basin is small as it was shown by calculations). We have got

*Proposition 4. The map* (2),(5) *with parameter values on the boundary* **3** *possesses a stable 3-point cycle.*

The results of computations of Sarkovskii boundary are given in Table 2. The 3-period boundary γ*(a) lies under the *flip* boundary for γ>0 and from the right for γ<0. It has the maximum γ*≈0.0454 at the point a=4.5 and the asymptote at a≈1.5. Increasing a and moving along the line γ=const with γ<0.0454, we cross over the 3-*period* boundary and can reach parameter domain IV or VI.

Behavior of the map G with parameter values in domain **IV** can be very complicated and its detailed study goes beyond the scope of this paper. We show only below that inside the domain **IV** there exists a sub-domain **IVa** such that the population gets to extinction at any initial size with the parameter values, belonging to this sub-domain.

### 5. 0-attracting domain

The equilibrium $N=0$ of the model corresponds to the extinction of a population; this point is stable (attractive) for any positive $\gamma$. Additionally, the equilibrium $N=0$ is unique and globally stable for the map $G(N)$ if the parametric point $(a,\gamma)$ belongs to domain **I** (see Fig.3). Otherwise, the map $G(N)$ has three fixed points: $N=0$, $N_1$, $N_2$, where $N_1$, $N_2$ are roots of (9), $0 < N_1 < N_2$. In the domains **II** and **III**, $N=0$ pertains a natural basin $[0, N_1)$; it can be shown that this basin enlarge when the parameter a increases, (see Fig.1a, 2a for domain II and Fig.1b, 2b for domain III). In the domain **IV** both fixed points $N=N_1$ and $N= N_2$ are unstable. Let us show that there exists sub-domain **IVa** of domain **IV** such that for parametric points $(a, \gamma) \in$ **IVa** the zero fixed point of $G$ is *globally* attractive, that is the fixed point $N=0$ attracts trajectories with almost all initial values (see Fig. **5a,b**).

Let $0 < \gamma < 1/8$ and $4a\gamma < 1$. Consider the curve **4** $\{a,\gamma\}$ given by the formula:

$$G^2(N^-) = N_1 \qquad (12)$$

where $N^-=(1-\sqrt{(1-8\gamma)})/2$ is the smallest root of equation (6), $N_1=(1-\sqrt{(1-4a\gamma)})/2a$ is the smallest root of equation (9). Let the point $(a_0,\gamma_0)$ belongs to this line and satisfies the condition:

$$G(N^-) = N^+ \qquad (13)$$

where $N^+=(1+\sqrt{(1-8\gamma)})/2$ is the largest root of (6). Recall that $N^-$, $N^+$ are respectively the maximum and minimum of the map $G$, and $N^- < N_1 < N^+$ (see Fig.-s **5a,b**). Denote **IVa** the set of points $(a,\gamma)$ bounded by curve (12).

*Proposition* 5.

(*i*) *For each $a > a_0$ two values $\gamma'' > \gamma_d$ satisfy the equality* (12); *the curve* (12) *has two monotonically decreasing branches in parameter space* $(a, \gamma)$ *(see Fig.-s 3). The vertex of this curve is the point* $(a_0, \gamma_0) = (3.938756.., 0.0417503..)$.

(*ii*) *The domain IVa is located inside domain IV.*

(*iii*) *For any $(a,\gamma) \in$ IVa almost all trajectories of map G tend to $N=0$ at $t \to \infty$ (excluding those with initial values from a totally disconnected set).*

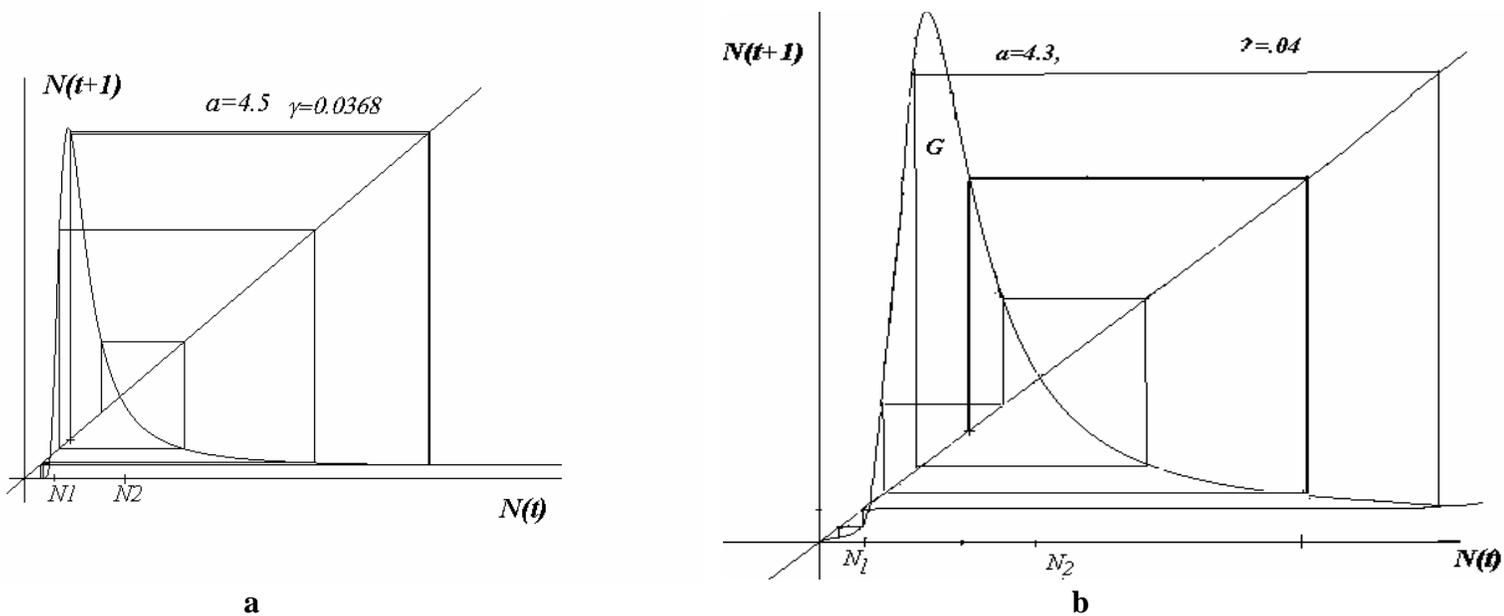

**a**                                        **b**

**Fig.5**. Phase portraits of model (1), (2), (5) in the domain IVa of the parameter portrait for
**a** - $a$= 4.5, $\gamma$= 0.0368, **b** - $a$= 4.3, $\gamma$=0.04.

The complete analytical proof of the Proposition based on the methods developed in (Singer, 1978, Devaney, 1989, et. al) and applied in (Thunberg, 2000) for analysis of the Ricker and Hassel population models, goes beyond the scope of this paper. The branches $\gamma=\gamma^u(a)$, $\gamma=\gamma_d(a)$ of the curve **4** calculated by the TRAX software program (Levitin, 1989) are given in Table 2.

Note that formula (12) was suggested in (Franke, Yakubu, 1994) who has analyzed a particular case of a map with hump and tail.

### 6. *Model* (**2**), (**5**) *with negative $\gamma$*

Map $G$ with *negative* $\gamma$ has a unique positive fixed point, which is the root $N_2 = (1+\sqrt{(1-4a\gamma)})/2a$ of equation (9) ("conjugate" root $N_1=(1-\sqrt{(1-4a\gamma)})/2a$ is negative). Analysis of the map $G$ given in Propositions 1-4 holds also for negative $\gamma$. For example, periodic and a-periodic trajectories exist in the model with any $\gamma<0$ and appropriate $a$. For better understanding the model dynamics under fixed $\gamma<0$ let us note, that when the parameter $a$ increases the fixed point $N_2$ decreases and tends to zero with $a\to\infty$; when $a$ crosses over the flip boundary, the fixed point $N_2$ becomes unstable. If $N_2$ is unstable, any orbit $\{N, G(N), G^2(N),...\}$ (with $N\neq N_2$) has infinite number of points that are less than $N_2$. This statement follows from Proposition 6 (see Appendix 2). We can conclude that a trajectory of the model with large *a* visits (infinitely often) a vicinity of origin bounded by $N_2$ and this vicinity is small if $N_2$ is small.

Let us return to the biological meaning of the model. It is necessary to distinguish formal characteristics of the solution of mathematical model and its interpretation. Recall that the solution of the model with *negative γ* and arbitrary non-zero initial value cannot vanish. So the event $N(t)$<<1 for some *t* should be interpreted as population extinction at time *t*. The model loses biological meaning for small *N* because the *growth rate for small values of N is unrealistically large* (tends to infinity as *N*→0). This can be a reason for the apparent chaoticity of the trajectories with very large amplitude, which visit a small neighborhood of the origin. It can be interpreted that the real population is very likely to go extinct. Therefore, in the framework of model (2), (5) *with negative γ* the appearance of periodic or a-periodic trajectories with large amplitude should be considered as a sign of either destruction of the population very soon or that the model with given parameter values may not be an adequate representation of population dynamics. Nevertheless, the model may describe the population dynamics outside of the vicinity of the origin.

**Remark.** The problem of biological adequacy of the model with negative *γ* arises for large values of *a*. To avoid this problem and to improve the model (2), (5) with negative *γ* it is possible to consider a truncated model, that is to replace the function *G*= *N*exp(*r*(*N*(*t*)) by the function *G*= *N*min[exp(*r*(*N*(*t*), *r*\*], where *r*\* is the maximal possible growth rate for a population. For large *r*\* we have observed the existence of a-periodic *bounded* trajectories. The studying of the truncated in detail is beyond the scope of this paper.

### 7. *Bifurcation portrait of the model.*

Phase-parameter portrait of the model given in Fig.3 allows us to predict the principal characteristics of the population dynamics of different kinds of rotifer species. The parameter space (*a* >0, *γ*) is dividing into eight domains of qualitatively different types of a population dynamics, characterizing by *Extinction*- I, IVa, *Monostability* –V, *Bistability* -II, *Periodic orbits* - III, VI, *Periodicity and Chaos* -IV, VII.

If variation in parameter *a* leads to crossing over the boundary *2* between domains II and III for *γ*>0 or between domains V and VI for *γ* <0, then the population dynamics is changed from oscillatory to steady-state and back, but it does not drive a population to extinction. So, boundary *2* can be considered *safe*.

Boundaries 1 and 4 are *dangerous* because their crossing over drives a population to extinction in domain **I** or in domain **IVa** correspondingly**.** Let us underline an essential difference of population behaviors close to these boundaries. If we start from a parametric point $\{a,\gamma\}$ in domain II then approaching to *1* with increase of the parameter *a* is accompanied by monotone closing together of stable and unstable equilibria $N = N_1$, $N = N_2$. In the boundary, $N_1 = N_2$ and after the crossing over the boundary both equilibriums disappear and the model has a single fixed point $N=0$. This way of the population extinction is well known.

Another way of the population extinction is realized when we start from a parametric point $\{a,\gamma\}$ in domain **IV.** Then approaching to the boundary *4* with increase of the parameter *a* is accompanied by *a*-periodic oscillations. These oscillations are "unexpectedly" terminated with crossing over the boundary at such a value of the parameter *a* that $G^2(N^-) = N_1$ where $N^-$ corresponds to the maximum of the function $G$ in (1) (see (12) and Proposition 5). Further increase of the parameter *a*, when a parametric point $\{a,\gamma\}$ enters into domain IVa where $G^2(N^-) < N_1$, goes the population to extinction.

At last, boundary *3* should be supposed also dangerous, at least for $\gamma < 0$, because its crossing over drives the model to domain **VII** where high-amplitude oscillations likely result the extinction of the population.

### 7. Dynamics of rotifer populations under toxicant exposure

In section 1, we considered examples of the dynamics of different natural rotifer populations. In the previous section we described the bifurcation portrait of the general model (2), (7). Let us now put the nine rotifer species whose parameter values are given in Table 1 into the phase-parametric portrait of Fig. 3.

Recall that parameter $\gamma$ characterizes the density-dependence of a population's dynamics, whereas parameter *a* characterizes the *quality* of an environment. The effects of toxicants on rotifer population dynamics is represented by parameter *a* because an increase in *a* increases population death rate (1). Using the parameter portrait (Fig.3), we can trace the development of different dynamical regimes as parameter *a* increases.

1) Models of population 1: *Keratella tropica*, 2: *Asplanchna girodi*, 4: *Brachionus dichotomus* and 3: *B. liratus* have $\gamma$-indexes $>1/16$. The parametric points given in Table 1 bring these species into domain II (Fig.3). Therefore, dynamics of these populations have a single non-trivial steady state. If parametric points of these species with growth of *a* move up to the boundary *1*, all species still remain in the domain II. This means that their population dynamics do not change qualitatively and there exists a single non-zero steady state. Equilibrium values of population size monotonically decrease with increasing *a* (see **Fig. 6a**). Populations become extinct when parameter *a* reaches a critical value, specific for each population, and equal to $a^* = 1/(4\gamma)$, which defines the boundary of domain I .

2) The model of population 6: *B. rotundiformis* has positive but small γ-*index*, $\gamma = 0.057 < \gamma^* = 1/16$. The parametric point belongs to domain II (Fig. 3). Therefore, the population exists in equilibrium steady state for small values of *a*. As *a* increases the parametric point $(a,\gamma)$ crosses overs the boundary of domain III and a periodical oscillation occurs in the model (Fig. 3). Further increase in *a* brings the parametric point $(a,\gamma)$ back to domain II thus to the steady state dynamics, but with smaller population size at equilibrium (**Fig. 6b**). Let us emphasize, that the line $\gamma = 0.057$ corresponding to γ-*index* of the model does not cross over the *3-period* boundary, whose maximal value is equal approximately 0.0454 (see Table 2). This means that a population will not go extinct before the value of parameter *a* reaches the boundary of domain I.

3) The model of population 5: *B. budapestinensis* has γ-index = 0.0637, which is very close to the boundary value $\gamma = 1/16 \approx 0.0624$. Thus, this model with $a = 2.925$ lies in domain II of the parameter portrait (Fig. 3), but is very close to the boundary between domains II and III. With small increases in *a,* the population enters domain I and eventually becomes to extinct with monotonically decreasing population size similar to species 1-4. Let us emphasize that placing the initial parametric point for population 5 almost at the boundary of domains II and III implies a specific dynamical behavior. Small changes in even one coefficient of the model (2), (3), (probably less than data measurement errors) may result in appearance of oscillatory dynamics.

4) The model of population 9: *Filinia pejleri* whose behavior for initial parameter values looks similar to that described above (Fig.1), actually has other dynamics. The model has a negative γ-index, $\gamma = -3.6625$ and lies in domain V of the phase-parametric portrait (Fig.3). As *a* increases and the parametric point reaches the boundary of domain VI, the steady state dynamics is changed to periodic oscillations and then to multi-period oscillations with growing amplitudes and eventually to an a-periodic oscillation after crossing over the *3-period* boundary and entering domain VII. Amplitudes of a-periodic oscillations increase dramatically if trajectories approach a small neighborhood near the zero point. Practically, this means that the population becomes extinct with chaotic oscillations of growing amplitude.

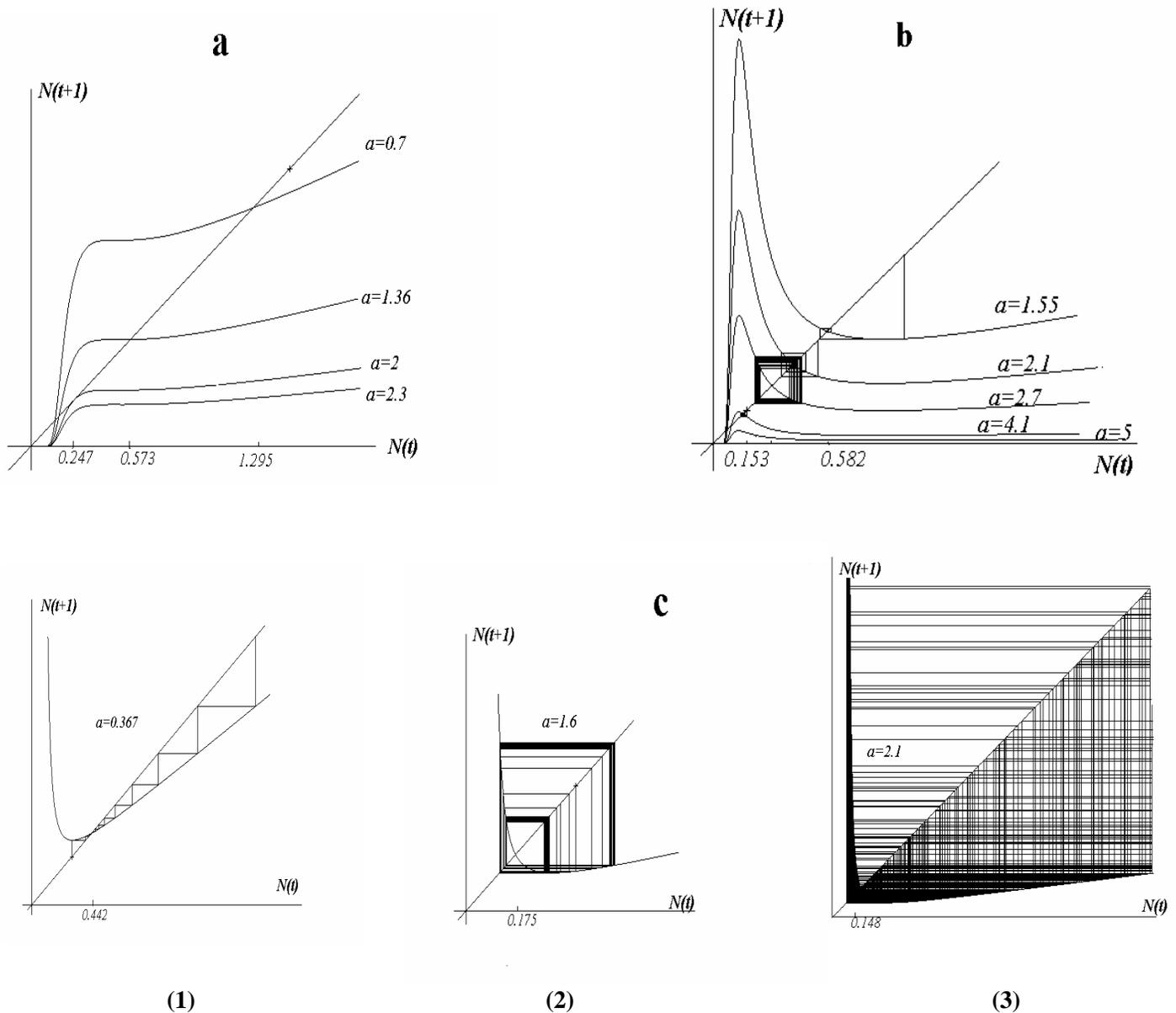

**(1)**                    **(2)**                    **(3)**

**Fig. 6**. **a** - Evolution of phase portrait of *B. dichotomus* ($\gamma$ =0.125):  with increasing toxicant exposure *a* steady equilibrium $N_2$ monotonically decrease (*a*=0.7, $N_2$=1.295; *a*=1.36, $N_2$= 0.573; *a*=2, $N_2$ =0.247); **b** -Evolution of phase portrait of *B. rotundiformis* ($\gamma$=0.057) with increasing toxicant exposure: stable steady state for *a*= 1.55 ($N_2$=0.582) and  *a*=2.1 ($N_2$ =0.4), stable oscillations for *a*=2.7, stable steady state for *a*=4.1 ($N_2$=0.153); **c** - Evolution of phase portrait of *Filinia pejleri* ($\gamma$=-3.6625) with increasing

*a*:  **(1)** stable steady state in domain V (*a*=0.367, $N_2$=0.442); **(2)** stable oscillations in the area VI (*a*=1.6, $N_2$=0.175); **(3)** multi-periodic oscillations in domain VII (*a*=2.1, $N_2$ =0.148) in phase space.

5) The same types of dynamics are observed in the models of populations 7: *Monostylla bulla* ($\gamma$ = -0.0539) and 8: *B. angularis* ($\gamma$ = -0.0638). Their paramteric points belong to domain VI. Note that the *M. bulla* parametric point is located just to the left of the domain VI boundary, whereas the *B. angularis* parametric point is located very close to the *3-period* boundary of domain VI. Thus non-periodic, chaotic oscillations with growing amplitude (Fig. 1e) are the typical dynamics of these models as parameter *a* increases.  The model of these population may be biologically unreliable when *a*>>1.

## 8. Discussion

This paper illustrates for the first time how two kinds of dynamical regimes - steady-state and oscillations (including complicated and even chaotic) - appear in *deterministic models* of natural rotifer populations observations in *stable environments*. It also illustrates how *small changes* in environmental quality can cause populations to transition to new dynamical regions, which increase the likelihood of the extinction.

Phase-parametric portraits of the models defined from equations (2), (3) (Fig.3) allow us to predict the principal characteristics of the population dynamics of other rotifer species not included in Table 1. For this problem, it is enough to calculate a species $\gamma$-index for given values of the environmental parameter *a* and place it in the parametrical portrait of Fig.3.

The principal contribution of this paper, therefore, is to define domains of parameter space where population persistence is possible. Using data from natural populations of nine rotifer species, we identify parameter *a* as a variable representing density-independent environmental conditions, so that an increase in *a* represents deterioration of environmental quality. Such deterioration is expected to occur with increasing exposure to toxicants, for example. Parameters *b* and *c* in our model represent density-dependent factors affecting population growth such as sensitivity to crowding or food limitation.

These parameters values have a solid empirical basis since all were derived from observations of natural rotifer populations. Time series analysis of short time series (<50 data points) of population abundance provided estimates of these values. Such data is readily collected, so there is wide application of the analytical techniques presented here. Moreover, the consensus model described the dynamics natural rotifer populations better than classical models like the logistic (Snell and Serra, 1998).

Five of the nine rotifer species investigated (*A. girodi, K. tropica, B. liratus, B. dichotomus* and *F. pejleri*) had population dynamical behaviors that lead to steady-states in stable environments, for the observed parameter *a* values. For the remaining four species, *M. bulla*, *B. budapestinensis, B. angularis* and *B. rotundiformis*, the parametric points lie near domain boundaries, so their dynamical behavior can change dramatically with small changes in the environmental quality parameter *a*. For example, *M. bulla* and *B. angularis* lie in domain VI which is characterized as periodic and a-periodic oscillations. Populations with those dynamics could persist for quite a long time as long as environmental quality remained stable. However, relatively small increases in parameter *a* will move *B. angularis* into domain VII where chaotic oscillatory dynamics is likely to lead to extinction. This example illustrates how toxicant exposure can drive a population to extinction not by directly increasing its mortality rate, but by changing the dynamical behavior of the population.

Therefore, the analysis of phase-parameter portrait shows three fundamentally different mechanisms of model population extinction. The *fir*st way is for stressors to increase death rates and decrease birth rates. Such a trend will eventually lead to population extinction (monotonically for representative point belonging to II or non-monotonically with crossing over III and I) if it is severe enough or persists long enough. This way can be realized, e.g., for the populations 1-6: *Keratella tropica*, *Asplanchna girodi, Brachionus dichotomus, B. liratus, B.budapestinensis, B.rotundiformis*, (see Fig.-s 6a,b).

The *second* possibility is subtler. In this case, small reductions in environmental quality due to toxicant exposure (increase in parameter *a*) can dramatically alter the dynamical behavior of a population. A population formerly in steady-state oscillation may enter a domain of chaotic dynamics, which mode increases the likelihood of population extinction. This way can be realized for populations 7-9: *7.Monostylla bulla, 8. B.angularis*, 9.*Filinia pejleri* with small increase in toxicant exposure**,** see Fig. 6c.

The *third* way is realized as a particular case of the second one. Under toxicant exposure a population may enter a domain of long/a-periodic oscillations and then to a domain where these oscillations are suddenly terminated and a population gets extinction for (almost) all initial sizes (domain IVa).

It seems prudent to explore how much toxicant exposure it takes thrust rotifer populations into a new dynamical regime. If these exposures are lower than those causing classical toxicity, this would have important implications for setting safe thresholds for toxicant exposures in the environment as well as ecological risk assessment.

**Acknowledgement**s

We would like to thank Dr. M. Borodovsky (Georgia Tech) for assistance and valuable discussions of the biological and mathematical problems. We appreciate to Dr. A.-A. Yakubu (Howard University) for  assistance with some computer simulations and proofs of results. The work of G.K. was partially supported by the grant from the National Institutes of Health to M.B.

**Table 1.** Coefficients *a, b, c, γ* of models (2), (3), (5)

| | Taxon | $N(t+1) = N(t)\exp(r(N(t))$ $r(N) = -a+b/N\ -c/N^2 \xrightarrow{N\to N/b}$ $r(N) = -a+1/N-\gamma/N^2$ | | | |
|---|---|---|---|---|---|
| | | *a* | *b* | *c* | *γ= c/ b²* |
| 1 | *Keratella tropica* | 0.547 | 0.511 | 0.050 | 0.19 |
| 2 | *Asplanchna girodi* | 0.305 | 0.081 | 0.001 | 0.152 |
| 3 | *B.dichotomus* | 1.359 | 1.344 | 0.225 | 0.125 |
| 4 | *Brachionus lyratus* | 1.425 | 1.12 | 0.095 | 0.076 |
| 5 | *B.budapestinensis* | 2.925 | 3.533 | 0.789 | 0.0632 |
| 6 | *B.rotundiformis* | 1.552 | 0.44 | 0.011 | 0.057 |
| 7 | *Monostylla bulla* | 1.865 | 1.422 | - 0.109 | - 0.0539 |
| 8 | *B.angularis* | 2.687 | 3.538 | - 0.799 | - 0.0638 |
| 9 | *Filinia pejleri* | 0.367 | 0.092 | - 0.031 | - 3.6625 |

**Table 2.** Parameter boundaries **3**($\gamma^*$) and **4**($\gamma^u$, $\gamma_d$) of model (2),(5)

| *a* | 1.6 | 1.8 | 1.85 | 2.0 | 2.3 | 2.7 | 3.1 | 3.3 | 3.5 |
|---|---|---|---|---|---|---|---|---|---|
| *γ* | -160.0 | -5.400 | -3.663 | -1.390 | -0.355 | -0.072 | -0.006 | 0.013 | 0.020 |

| *a* | 3.94 | 3.95 | 4.0 | 4.3 | 4.5 | 5 | 5.5 | 6 | 6.5 |
|---|---|---|---|---|---|---|---|---|---|
| $\gamma^*$ | 0.0450 | 0.0452 | 0.0455 | 0.0459 | 0.0454 | 0.0431 | 0.0405 | 0.0381 | 0.0357 |
| $\gamma^u$ | 0.04215 | 0.0428 | 0.0440 | 0.0450 | 0.0446 | 0.0426 | 0.0402 | 0.0378 | 0.0355 |
| $\gamma_d$ | 0.04213 | 0.0405 | 0.0386 | 0.0335 | 0.0312 | 0.0272 | 0.0243 | 0.0220 | 0.0202 |

## Appendix 1

*Proof of Proposition* 1.

Critical points $N^+=(1+v)/2$ , $N^-=(1-v)/2$, where $v=\sqrt{(1-8\gamma)}$ and $\gamma<1/8$ are the roots of

$$G_N(N)=\ \exp(r(N))\ (1-1/N+2\gamma/N^2)= \exp(r(N))(-\ r(N)\ +1-a+\gamma/N^2). \qquad (A1)$$

$G$ has maximum at $N^-$ and minimum at $N^+$ because $G_{NN}(N^{\pm})= \pm e^{r(N\pm)}v/(N^{\pm})^2$ .

Positiveness of $SG(N)\equiv G_{NNN}(N)/G_N(N)-3/2(G_{NN}(N)/G_N(N))^2 =$

$16\gamma^4/N^{10} -32(1-N)/\ N^9 + 12\gamma^2(2-4N+3N^2)/\ N^8 -4\gamma(2-6N+\ 6N^2+\ N^3)/\ N^7+ (1-4N+6N^2)/N^6$

was checked by computations.

*Proof of Proposition* 2.

Let $N*$ be a fixed point of $G$: $N* = G(N*)$. $N*$ is *attracting* if $|\mu(N*)| = |G_N(N*)| < 1$.

1) The stability $N* = 0$ for $\gamma > 0$ and instability for $\gamma \le 0$ are geometrically evident.

2) If a fixed point $N* \ne 0$, then $r(N*) = 0$. Due to (A1), the condition of stability of $N*$ is

$$|1 - 1/N* + 2\gamma/N*^2| < 1 \qquad (A2)$$

or, equivalently,

$$|1 - a + \gamma/N*^2| < 1. \qquad (A2')$$

At the boundary **1** equation (4) has only double root $N* = 1/(2a)$, and $G_N(N*) = 1$ due to (A1). Thus $N*$ is unstable. Consider now $\gamma < 1/(4a)$. Here $G$ has fixed points $N_1 < N_2$. The condition (A2) can be written as the system of inequalities:

$$N* > 2\gamma \qquad (A3)$$
$$2 - 1/N* + 2\gamma/N*^2 > 0.$$

Let us show that $N_1$ does not satisfy even the first inequality (A3) for any positive $\gamma$ and $a$ whereas $N_2$ satisfies both inequalities (A3) for certain $\gamma$, $a$. Denote $u = \sqrt{1 - 4a\gamma} > 0$. Then $N_1 > 2\gamma \Leftrightarrow [1 - \sqrt{(1 - 4a\gamma)}] > 4a\gamma \Leftrightarrow (1 - 4a\gamma) - \sqrt{(1 - 4a\gamma)} > 0 \Leftrightarrow u^2 - u > 0$, Thus, $N_1 > 2\gamma$ only for $u > 1$, that is if $a\gamma < 0$; it contradicts to the supposition that $a$, $\gamma$ and $N_1$ are positive.

Next, $N_2 > 2\gamma \Leftrightarrow u^2 + u > 0$. Consider the second inequality for $N_2$. From (A2'): $2 - 1/N_2 + 2\gamma/N_2^2 > 0 \Leftrightarrow 4a^2\gamma/(1+u)^2 > a - 2 \Leftrightarrow (1 - u^2)a/(1+u)^2 > a - 2 \Leftrightarrow a(1-u) > (a-2)(1+u) \Leftrightarrow u(a-1) < 1$. If $a \le 1$, then the last inequality is true for any $\gamma$ such that $4a\gamma < 1$; if $a > 1$ then it is true for $4a\gamma < 1$ and $\gamma > (a-2)/(4(a-1)^2)$.

*Proof of Proposition* 3

It is enough to check that $F_1(N_2)F_2(N_2) \ne 0$ where $F_1(N_2) = G_{Na}(N_2;\gamma,a)$ and $F_2(N_2) = 1/2(G_{NN}(N_2;\gamma,a))^2 + 1/3 G_{NNN}(N_2;\gamma,a)$, for $(a, \gamma)$ located at bounary **3**: $\gamma = (a-2)/(4(a-1)^2)$ (Kuznetsov, 1997, Th. 4.3). Indeed, $F_1(N_2) = \exp(r(N_2))(-r(N_2) + 1 - a + \gamma/N_2^2)r_a(N_2) = -(1 - a + \gamma/N_2^2) = 1$, because of $r_a(N) = -1$ and $\gamma/N_2^2 = a - 2$. As $N_2 = 1/(2(a-1))$, $F_2(N_2) = 8/3(a-1)^2(3a^2 - 15a + 20) > 0$ for $a > 1$.

**Appendix 2**

**Proposition 6.** *Let a function $G(x)$ is continuously differentiable and the map $x_{n+1} = G(x_n)$ has a fixed point $x*$ such that $G_x(x*) < 0$ and $G(x) < x$ for all $x > x*$. Then any orbit {x,* $G(x)$, $G^2(x),...$ *(with $x \neq x^*$) has at least one point less than $x^*$, i.e. $G^k(x) < x^*$ for some k.*

*If, additionally, $G(x) > x$ for all $x < x^*$, then any orbit $\{x, G(x), G^2(x),...\}$ (with $x \neq x^*$) has infinite number of points less than $x^*$, i.e. $G^k(x) < x^*$ for infinite number of k, and infinite number of points more than $x^*$, i.e. $G^s(x) > x^*$ for infinite number of s.*

*Proof.* Let us denote $x_n = G^n(x)$, $x_0 = x$. If $x > x^*$, then $x_1 = G(x) < x_0$ by conditions. Again, if $x_1 > x^*$, then $x_2 = G(x_1) < x_1$. So either for some $k$ $x_k = G^k(x) < x^*$ (as desired), or the orbit $\{x, x_1, x_2,...\}$ is monotonically decreasing sequence and $x_k > x^*$ for all $k$. Then there exist $\lim_{k \to \infty} x_k = x^{**} \geq x^*$. Evidently, $G(x^{**}) = x^{**}$, so $x^{**} = x^*$ (by conditions, if $x^{**} > x^*$, then $G(x^{**}) < x^{**}$). Thus we have a consequence $\{x_k > x^*\}$ which tends to $x^*$ and $G(x_k) > x^*$ *for all k*. But it contradicts to the condition $G_x(x^*) < 0$. The first part of the proposition is proved.

Let $G(x) > x$ for all $x < x^*$. It has been proved that there exists $k$ such that $x_k < x^*$. By the same way we can prove that either there exist such $s$ that $x_{k+s} > x^*$, or the sequence $x_{k+1}$, $x_{k+1},...$ is monotonically increasing, tends to $x^*$ and $G(x_{k+s}) < x^*$ for all $s$. But it again contradicts to the condition $G_x(x^*) < 0$. Repeating both steps, we complete the proof.